\documentclass[twocolumn]{article}  % Using article class instead of elsarticle
\usepackage[utf8]{inputenc}
\usepackage[english]{babel}
\usepackage{times}
\usepackage{lineno}
\usepackage{amsmath}
\usepackage{pdfpages,multirow,ragged2e}
\usepackage{hyperref}
\usepackage{siunitx}

\usepackage[
sorting=none
]{biblatex}

\newcommand\scalemath[2]{\scalebox{#1}{\mbox{\ensuremath{\displaystyle #2}}}}

\begin{document}

% FRONTMATTER
\title{A wireless method to obtain the longitudinal beam impedance from scattering parameters}
\author{Chiara Antuono$^1$, Carlo Zannini$^1$, Andrea Mostacci$^2$, Mauro Migliorati$^2$ \\
\small $^1$CERN, Geneva, Switzerland \\
\small $^2$University of Rome "La Sapienza", Rome, Italy}
\date{April 2021}
\maketitle
\begin{abstract}
The coaxial wire method is a common and appreciated technique to assess the beam coupling impedance of an accelerator element from scattering parameters. Nevertheless, the results obtained from wire measurements could be inaccurate due to the presence of the stretched conductive wire that artificially creates the conditions for the propagation of a Transverse ElectroMagnetic (TEM) mode.

The aim of this work is to establish a solid technique to obtain the beam coupling impedance from electromagnetic simulations, without modifications of the device under test. In this framework, we identified a new relation to get the longitudinal resistive wall beam coupling impedance of a circular chamber directly from the scattering parameters and demonstrated that it reduces to the exact theoretical expression. Furthermore, a possible generalization of the method to arbitrary cross-section chamber geometries has been studied and validated with numerical simulations. 
\end{abstract}

\section{Introduction}
The beam coupling impedance describes the electromagnetic interaction between the particle beam and the accelerating structure.
Ideally, the beam coupling impedance of a device should be evaluated by
exciting the device with the beam itself. However, in most cases, this solution
is not possible, and one must resort to alternative methods to consider the
effect of the beam.

A well-established technique consists in approximating the beam by a current pulse flowing through a wire stretched along the beam axis, resulting in the development of the stretched Wire Method (WM)~\cite{Faltens,SandsRees,Pedersen,VaccaroWire}. Nevertheless, the results obtained with wire measurements might not entirely represent the solution of our initial problem, because the presence of the stretched wire perturbs the EM boundary conditions. The most evident consequence of the presence of another conductive medium in the center of the device under study is the artificial propagation of the TEM mode through the device, with zero cut-off frequency. The presence of a TEM mode among the solutions of the EM problem will have the undesired effect of causing additional losses.
In this regard, attention has been focused on possible approaches without modification of the Device Under Test (DUT). Wireless measurements have already been proposed in~\cite{Lambertson} and performed in~\cite{TMmeas} above the cut-off frequency of the device under test (DUT), where an approximated formula has been employed.

An exact formula to obtain the longitudinal beam coupling impedance of the accelerator components is presented in this paper.
The new formula, relating the longitudinal beam coupling impedance and the scattering parameters, has been analytically validated for a resistive circular chamber also below its cut-off frequency. Furthermore, a possible generalization to arbitrary chamber shapes above the cut-off frequency has been explored.

\section{Wire method} \label{wireM}
\begin{figure}[!htb]
   \centering
   \includegraphics*[width=1\columnwidth]{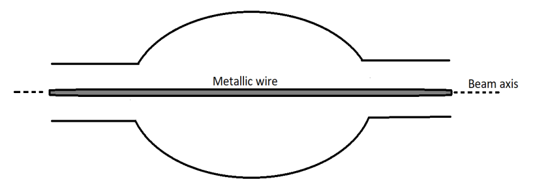}
   \caption{Schematic view of the wire measurement.}
   \label{fig:WireSetup.png}
\end{figure} 
The excitation produced by a relativistic beam in a device can be simulated by placing a conductor along the axis of the device. The idea behind this method relies on replacing the current pulse produced by a beam with a current pulse having the same temporal behavior but flowing through a wire stretched along the beam axis. As a result, the DUT is transformed into a coaxial transmission line allowing the propagation of TEM mode.
%The excitation produced by a relativistic beam in a device can be simulated by placing a conductor along the axis of the device. Therefore a metal wire is stretched in the DUT which is then transformed into a coaxial transmission line allowing the propagation of a TEM mode. Thus the longitudinal coupling impedance can be inferred from the properties of such a transmission line, provided that only the (fundamental)
%TEM mode is propagating.%is the characteristic impedance formed by the reference pipe with the axial wire as a coaxial transmission line structure;
In general, any transmission line can be characterized by measuring its scattering S-parameters and, in particular, it has been demonstrated that the beam coupling impedance can be related to the transmission scattering parameter. A number of approximated formulae were derived and can be found in the literature (e.g. see~\cite{Faltens,SandsRees,Pedersen}), depending on the characteristics of the DUT and on the kind of approximation made.
In 1993, V. Vaccaro derived a more rigorous and accurate formula to calculate the beam coupling impedance, based on the transmission line theory~\cite{VaccaroWire}.
He demonstrated that the longitudinal beam coupling impedance can be computed as follows:
\begin{equation}
    Z= Z_{c}\ln{\frac{S^{DUT}_{21}}{S^{REF}_{21}}} (1+\frac{\ln{S^{DUT}_{21}}}{\ln{S^{REF}_{21}}}),
    \label{eqWMGen}
\end{equation}
where $Z_{c}$ is the characteristic impedance of the coaxial transmission line structure formed by the DUT with the stretched wire, as shown in Fig. \ref{fig:WireSetup.png}.
The transmission scattering parameter $S^{DUT}_{21}$ refers to the 2-Port DUT, while the $S^{REF}_{21}$ refers to the related reference structure (smooth reference beam pipe).
If the $S^{DUT}_{21}$ is close to the $S^{REF}_{21}$, which is usually the case for many accelerator components, Eq.~(\ref{eqWMGen}) can be approximated by:
\begin{equation}
   Z\approx 2 \cdot Z_{c}\ln{\frac{S^{DUT}_{21}}{S^{REF}_{21}}}.
    \label{eqWM}
\end{equation}
However, it has been proven that this method provides inaccurate results below cut-off (\cite{VaccaroWireLimitations2,VaccaroWireLimitations}). As already mentioned, this behavior is due to the presence of the wire that introduces a TEM wave, which intrinsically has a zero cut-off frequency. All the resonant frequencies are depleted because of the power drained in the pipes by the TEM mode. This additional power loss drastically lowers the high Q resonances of the DUT. At frequencies above the cut-off, there are indications that this method gives fairly good results~\cite{VaccaroWireLimitations}.

\section{Wireless method} 
\label{newM}
\begin{figure}[!htb]
   \centering
   \includegraphics*[width=1\columnwidth]{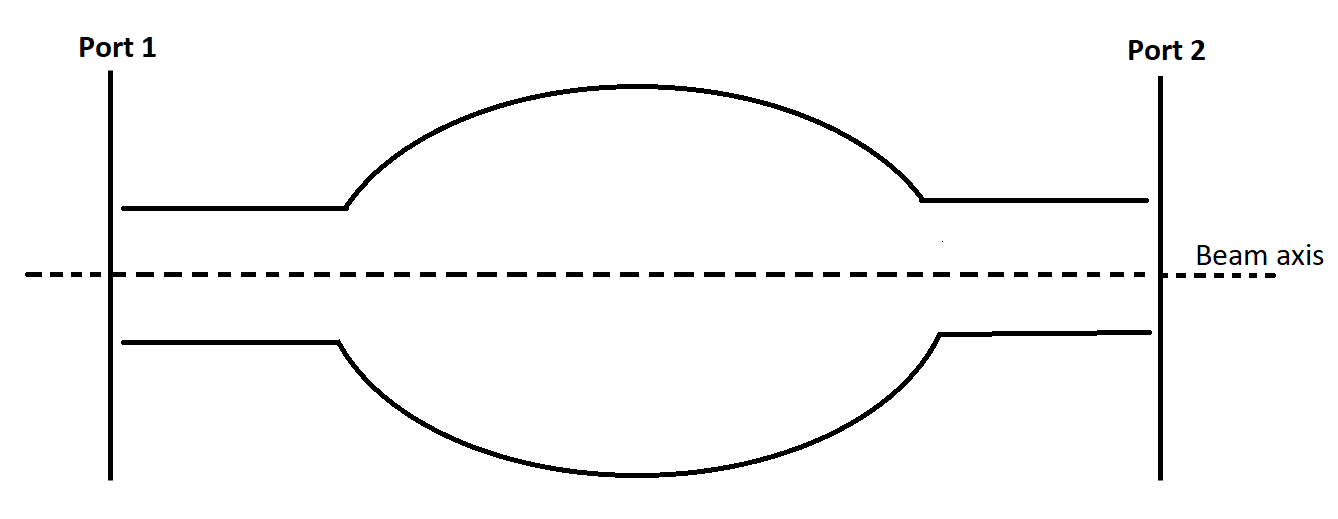}
   \caption{Schematic view of the wireless measurement. }
   \label{fig:PipeNoWire.png}
\end{figure} 
The longitudinal beam coupling impedance is related to the energy loss of the electromagnetic wave propagating in the structure and, therefore, is intrinsically linked to the transmission scattering parameter. 
Given these considerations and by analogy with the WM, we sought for a Log-formula similar to Eq.~(\ref{eqWM}) to express the beam coupling impedance by using the first propagating Transverse Magnetic (TM) mode in the DUT.
    % \begin{figure}[h!]
    % \centering
    % \includegraphics[width=1\linewidth,height=0.17\textheight]{Port_RW}
    % \caption{Schematic view of a section of a resistive wall chamber}
    % \label{fig:Port_RW}
    % \end{figure}\\
The proposed relation to evaluate the impedance, without modifications of
the DUT, has the following form:
\begin{equation}
    Z= -K \cdot Z_{mode}\ln{\frac{S^{DUT}_{21}}{S^{REF}_{21}}}.
    \label{eq}
\end{equation}

The $Z_{mode}$ is the characteristic wave impedance of the TM propagating mode. %The transmission scattering parameter $S^{DUT}_{21}$ refers to the 2-Port DUT, that is the chamber with finite electric conductive walls, while the $S^{REF}_{21}$ refers to the related reference structure, in this case, the chamber with Perfect Electric Conductive (PEC) walls. %The reference scattering parameter has been involved for normalization and in order to obtain an accurate evaluation of the impedance even under the cut-off frequency of the pipe. 
The term $K$ is a constant to be determined in the analytical derivation.

\subsection{Analytical validation for a circular beam chamber in the classical thick wall regime} \label{newMCalculation}
The proposed formula has been analytically validated for the simple case of a circular resistive wall chamber of radius $b$, wall conductivity $\sigma$, and length $L$, both below and above the cut-off frequency of the chamber. The longitudinal beam coupling impedance of the circular resistive chamber can be analytically calculated by using the following well-known equation~\cite{Chao}:
\begin{equation}
    Z^{theory}=\frac{\zeta_{s}}{2\pi b}L,
    \label{Theory}
\end{equation}
%by placing the constant term $K$ equal to $\frac{1}{2\pi}$
where, $\zeta_{s}=\zeta_{R}+j\zeta_{J}$ is the equivalent surface impedance of the wall. In the classical thick wall regime, $\zeta_{s}=\zeta(1+j)=\sqrt{\frac{\omega \mu_{0}}{2\sigma}}(1+j)$, $\omega$ is the angular frequency and $\mu_{0}$ the permeability of free space.
In order to validate the proposed approach we demonstrate that the formula of Eq.~(\ref{eq}) reduces to Eq.~(\ref{Theory}). The analytical expression of the $|S_{21}|$ of the resistive circular pipe is derived from the attenuation constant $\alpha$ by imposing the conservation of the energy. It turns out to be the following:
\begin{center}
\begin{equation}
    |S_{21}|=e^{-|\alpha|z}.
    \label{S21fromalfa}
\end{equation}
\end{center}

The attenuation constant $\alpha$ for a lossy circular pipe is obtained in~\cite{alphaArt} applying the Leontovich boundary condition and can be written as follows:
\begin{footnotesize}
\begin{equation}
    \alpha=\Im{\Big[\sqrt{k_{0}^2-\frac{1}{b^2}\Big[u_{nm}+\frac{jk_{0}^2}{\omega(\frac{u_{nm}}{b})(\frac{\mu_{0}}{\zeta_{s}}+\epsilon_{0}\zeta_{s})}\Big]^2}\Big]}.
    \label{alfa}
\end{equation}
\end{footnotesize}
The terms $k_{0}$, $\epsilon_{0}$ are the wave number and the permittivity of free space, and $u_{nm}$ is the $m^{th}$ zero of the Bessel function $J_{n}$. In the upcoming computation, only the first TM propagating mode is considered, which is denoted as $TM_{01}$ with the values $m=0$ and $n=1$. For the sake of simplicity, the subscripts are omitted everywhere.

\begin{figure}[!htb]
   \centering
   \includegraphics*[width=1\columnwidth]{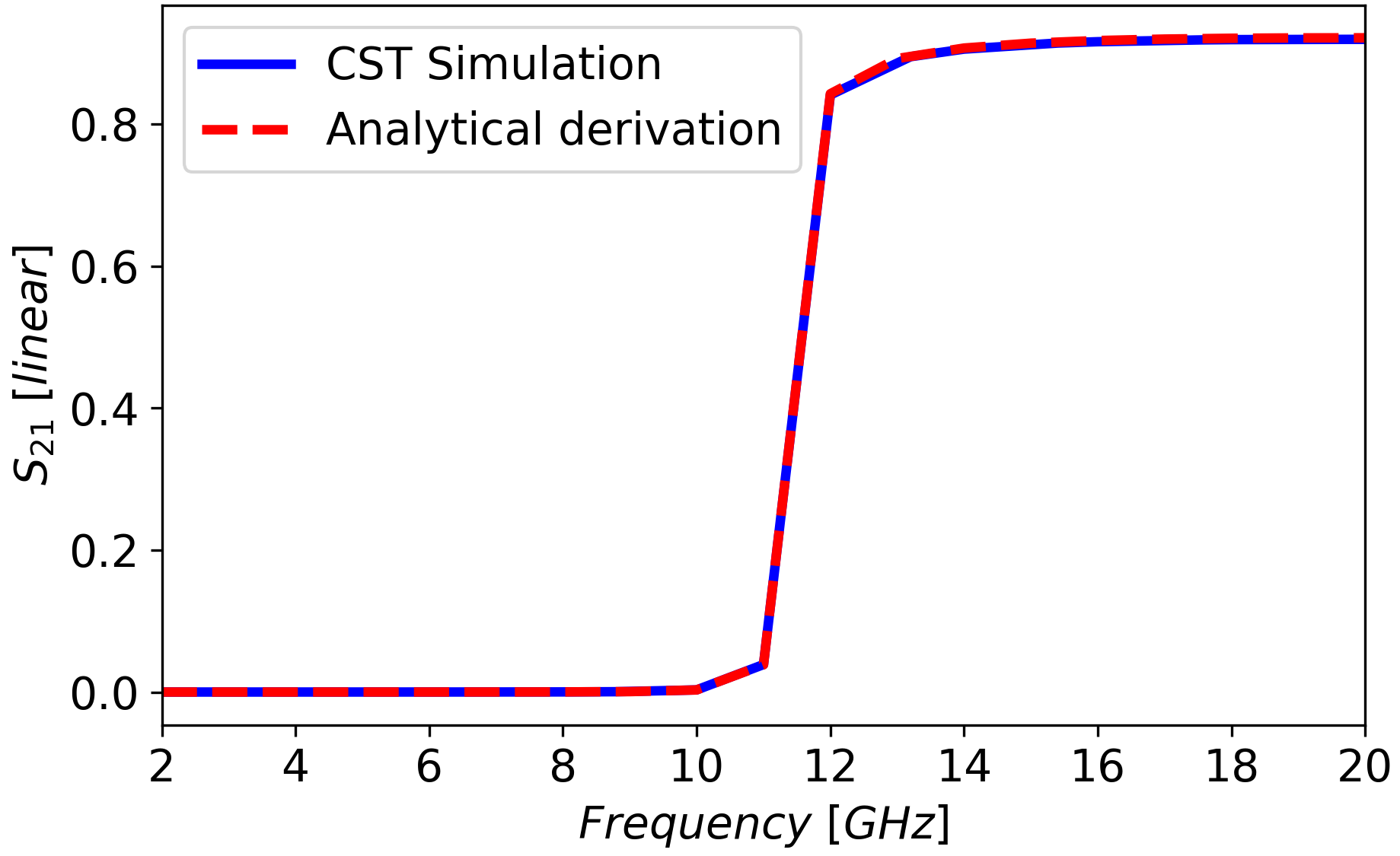}
   \caption{Comparison between the $|S_{21}|$ from the analytical computation and CST simulation for a circular chamber with $b=\SI{10}{mm}$, $\sigma=\SI{3000}{S/m}$ and $L=\SI{50}{mm}$. }
   \label{fig:S21Conf.png}
\end{figure}

Figure \ref{fig:S21Conf.png} shows the perfect agreement that has been reached between the simulated $|S_{21}|$ for a given resistive chamber and the analytical expression of $|S_{21}|$ computed with Eqs.~(\ref{S21fromalfa}),~(\ref{alfa}). 
Eqs.~(\ref{S21fromalfa}),~(\ref{alfa}) can be substituted in Eq.~(\ref{eq}), where:
\begin{center}
$Z_{mode}=Z_{TM}=\frac{\sqrt{k_{0}^{2}-\frac{u^{2}}{b^2}}}{\omega\epsilon_{0}}$,
\end{center}
and the following expressions of the longitudinal impedance are obtained below and above the chamber cut-off frequency~\cite{ThesisChiara}:
\begin{equation}
Z_{{{below}}}=-K\cdot Z_{TM}(\sqrt{T_{1}}-\sqrt[4]{T^{2}_{2}+T^{2}_{3}}) L
\label{Zbe}
\end{equation}
\begin{equation}
Z_{{{above}}}=-K\cdot Z_{TM} \Big(\sqrt[4]{T^{2}_{2}+T^{2}_{3}}\cdot \frac{-T_{3}}{2T_{2}}\Big) L,
\label{Zab}
\end{equation}
where $T_{1}=T_{2}-T_{3}$, $T_{2}=k_{0}^2-\frac{u^2}{b^2}+2\omega\epsilon_{0}\frac{\zeta}{b}$ and $T_{3}=2\omega\epsilon_{0}\frac{\zeta}{b}$.
Under the assumption that the structure can be treated as a planar geometry, which means that the radius of curvature is much greater than the skin depth ($b >> \delta=\sqrt{\frac{2}{\omega\mu_{0}\sigma}}$),
it can be shown that, both below and above the chamber cut-off frequency, the longitudinal impedance of Eq.~(\ref{eq}) %with the substitution of the analytical expressions of the above mentioned relevant parameters, 
reduces to Eq.~(\ref{Theory}), by placing the constant term $K$ equal to $\frac{1}{2\pi}$~\cite{ThesisChiara}.

\begin{figure}[!htb]
   \centering
   \includegraphics*[width=1\columnwidth]{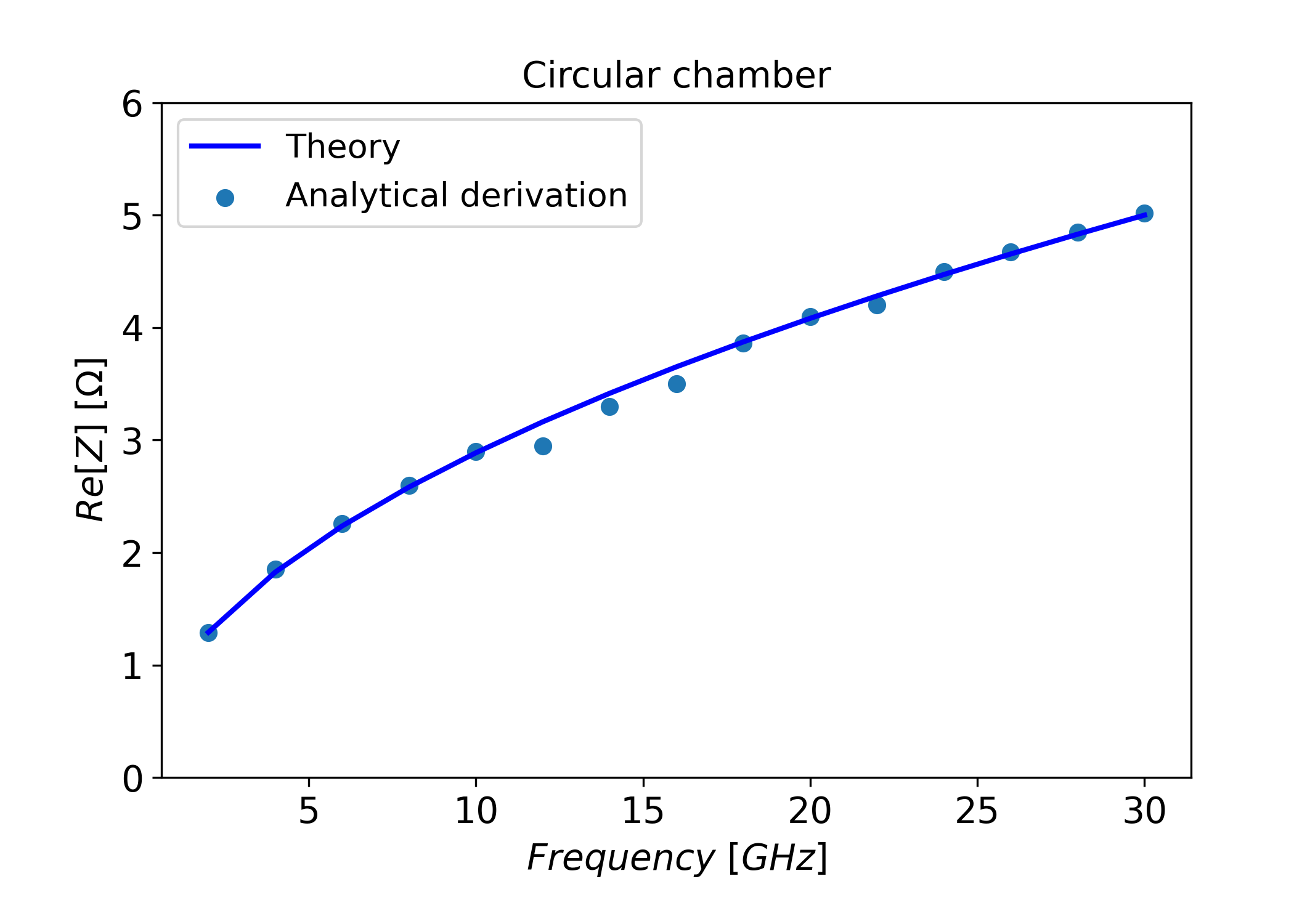}
   \caption{Comparison between the longitudinal impedance of Eqs.~(\ref{Zbe}),~(\ref{Zab}) and Eq.~(\ref{Theory}). The agreement is not perfect around $11$ GHz, the cut-off frequency of the chamber, due to the analytical approach which does not provide an estimation of the impedance at the cut-off frequency.}
   \label{fig:PlotCircularImp.png}
\end{figure}

This proves the correctness of the relation between the scattering parameter $S_{21}$ and the longitudinal beam coupling impedance proposed in Eq.~(\ref{eq}) that becomes:
\begin{equation}
    \Re{(Z)}= -\frac{1}{2\pi} \cdot Z_{TM} \ln{\frac{|S^{DUT}_{21}|}{|S^{REF}_{21}|}},
   \label{eqCirRe}
\end{equation}
where, in the classical thick wall regime, the real and imaginary part are the same.
Figure~\ref{fig:PlotCircularImp.png} displays the comparison between the impedance from the analytical derivation and the theoretical one.

\section{Generalization of the wireless method to arbitrary cross-section geometries}
The generalization of the method in~\ref{newMCalculation} to arbitrary shapes of the vacuum chambers has been explored. As a first step, the case of the rectangular chamber has been studied to investigate the potential of the method with non-axially symmetric structures. 
The result is that, above the cut-off frequency of the chamber, the impedance can be derived from the scattering parameter $S_{21}$ with the following relation~\cite{ThesisChiara}:
\begin{equation}
     \Re{(Z)}= -\frac{G\cdot F}{2\pi} \cdot Z_{TM} \ln{\frac{|S^{DUT}_{21}|}{|S^{REF}_{21}|}},
    \label{eqGen}
\end{equation} 
where $F$ and $G$ are geometrical factors. In particular, $F$ is the longitudinal form factor for the rectangular chamber (see~\cite{Yokoya,ThesisCarlo}). Analytical expressions of $F$ exist for the rectangular and elliptical chambers and could be computed for any geometry with simulations since it is given by the ratio of the wake function of the chamber under test and the wake function of the reference circular chamber (see ~\cite{ThesisCarlo}):
\begin{center}
    $F=\frac{w(z)^{DUT}}{w(z)^{CIR}}.$
\end{center}

\begin{figure}[!tbh]
     \centering
     \includegraphics[width=.80\columnwidth]{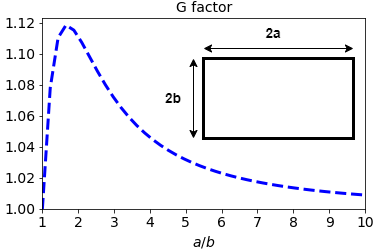}
     \caption{G versus $a/b$.}
     \label{fig:Gfac1.png}
\end{figure}

The theoretical expression of the beam coupling impedance (Eq.~(\ref{Theory})) can be generalized to arbitrary cross-section chambers using the form factor $F$.
The $G$ factor is related only to the maximum half-width $a$ and half-height $b$ of the cross-section of the DUT as defined by the following expression~\cite{ThesisChiara}:
\begin{center}
    $G=\frac{(b^2+a^2)a}{b^3+a^3}.$
\end{center}
Its behaviour versus the aspect ratio of the chamber is displayed in Fig.~\ref{fig:Gfac1.png}. 
     
It is worth mentioning that Eq.~(\ref{eqGen}) is a more general expression of Eq.~(\ref{eqCirRe}). In fact, for circular chambers, $F$ and $G$ are equal to 1, and Eq.~(\ref{eqGen}) would reduce exactly to Eq.~(\ref{eqCirRe}).
The method has also been tested in simulations for resistive wall chambers with elliptical and octagonal cross-sections.
%For elliptical cross section, the longitudinal form factor F is already known (see \cite{Yokoya} ), and for the octagonal case is computed with respect to a reference circular chamber with CST simulations, as could be done for any kind of shapes (see~\cite{ThesisCarlo}, ~\cite{fOct} ).
These simulations suggest that Eq.~(\ref{eqGen}) is a general expression that could be applied to obtain the longitudinal beam coupling impedance of arbitrary cross-section chambers (see Section~\ref{Comp}).

\section{Beyond the thick wall regime: the analytical validation for the general case of complex surface impedance}
The analytical validation in~\ref{newMCalculation} has been carried out making some simplifications in favor of the ease and compactness of the treatment, as well as in the use of the formula to determine the impedance. The most relevant assumption is $\zeta_{s}=\zeta(1+j)=\sqrt{\frac{\omega \mu_{0}}{2\sigma}}(1+j)$, as it happens in the classical thick wall regime. In this section, we present a generalization of the applicability of the method, even beyond the thick wall regime.
In fact, by considering the most general case of complex surface impedance as $\zeta_{s}= \zeta_{R}+j\zeta_{J}$, and repeating all the calculations of~\ref{newMCalculation}, the validation can be carried out in the same way. 
Under the same assumption of planar geometry ($b >> \delta=\sqrt{\frac{2}{\omega\mu_{0}\sigma}}$), the longitudinal impedance of Eq.~(\ref{eq}) reduces to Eq.~(\ref{Theory}), if the following modifications are imposed (see Appendix~\ref{AppS21},~\ref{AppValidation}):
\begin{equation}
Z= -\frac{1}{2\pi M} \cdot Z_{mode}\ln{\frac{S_{21DUT}}{S_{21REF}}},
\label{eqMCA}
\end{equation} %by placing the term $K$ equal to $\frac{1}{2\pi M}$, where $M$ is a correction factor given by:\\
where M is a correction factor given by:

\begin{equation*}
M = \begin{cases}
M=\frac{\mu_{0}N_{1}}{uD} (u - \frac{\omega\mu_{0}\epsilon_{0}b}{u} \frac{\zeta_{J}N_{2}}{D} ), &\text{for $\Re({Z})$} \\
1, &\text{for $\Im({Z})$}\\
\end{cases}
\end{equation*}
where:
\begin{center}
$N_{1}=\mu_{0}+\epsilon_{0}(\zeta^{2}_{R}+\zeta^{2}_{J})$\\
$N_{2}=\mu_{0}-\epsilon_{0}(\zeta^{2}_{R}+\zeta^{2}_{J})$\\
$D=\mu^{2}_{0}+\epsilon^{2}_{0}(\zeta^{2}_{R}+\zeta^{2}_{J})^2+2\epsilon_{0}\mu_{0}(\zeta^{2}_{R}-\zeta^{2}_{J})$.\\
\end{center}

\begin{figure}[!tbh]
     \centering
     \includegraphics[width=1\columnwidth]{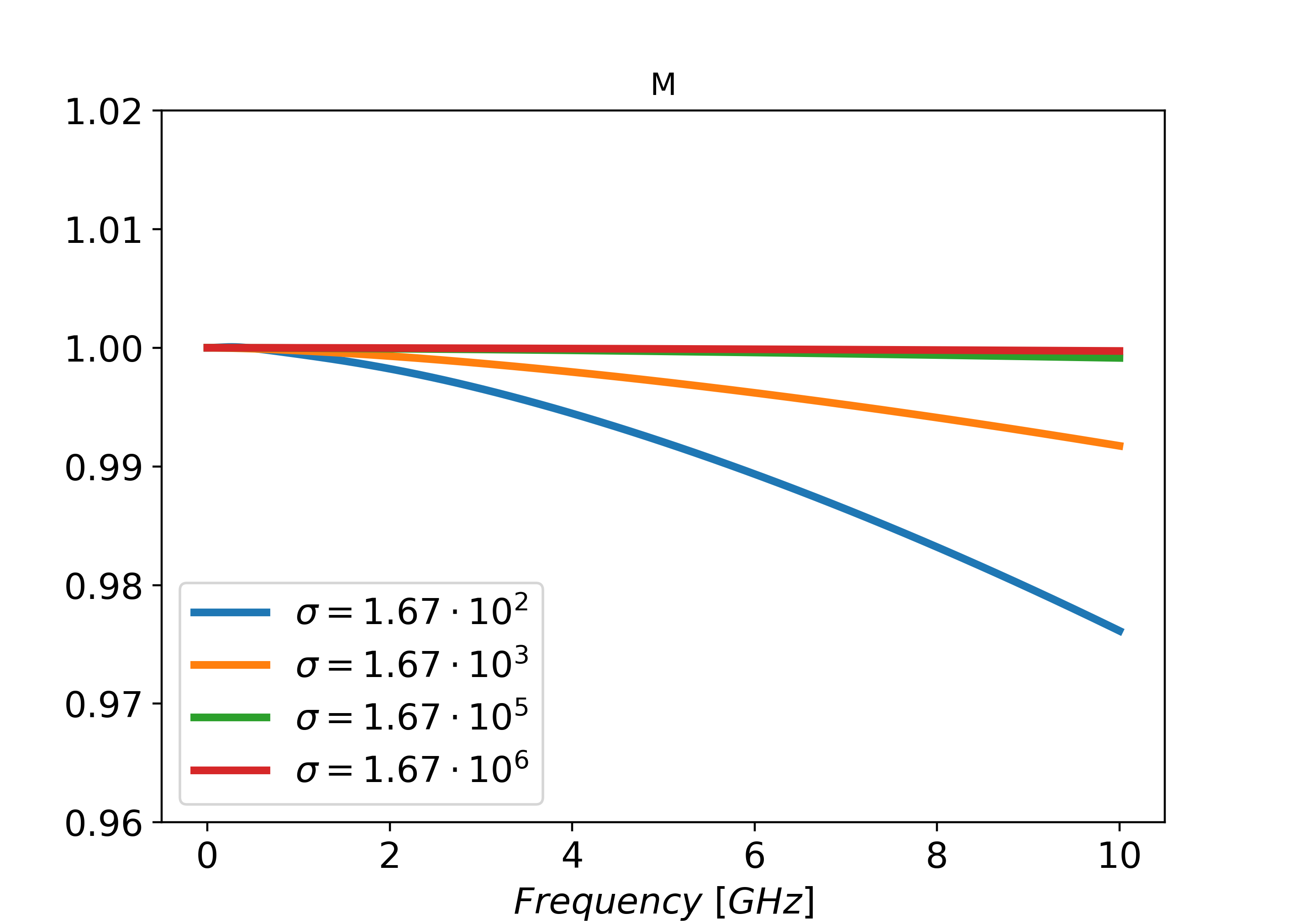}
     \caption{Correction factor M as a function of frequency and for different wall conductivities $\sigma$. It is obtained from the exact analytical validation in~\ref{AppValidation}.}
     \label{fig:FAKT.png}
\end{figure}

\begin{figure}[!tbh]
     \centering
     \includegraphics[width=1\columnwidth]{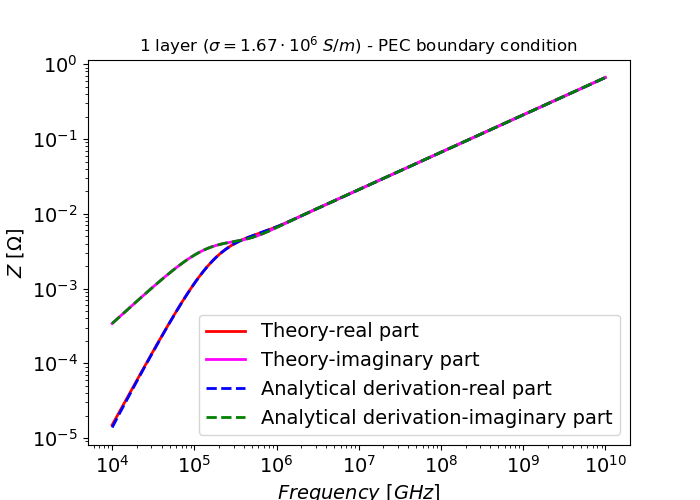}
     \caption{Longitudinal impedance for a circular chamber with $b=\SI{18.4}{mm}$, wall-thickness $s=\SI{1}{mm}$, conductivity $\SI{1.67e5}{S/m}$ and a PEC boundary condition.}
     \label{fig:BeyondThickWall.png}
\end{figure}

The correction factor found in the analytical treatment has the disadvantage of depending on the DUT material properties, since the real and imaginary parts of the surface impedance $\zeta_{R}$ and $\zeta_{J}$ are involved. The behavior as a function of the frequency and for different conductivities is shown in Fig.~\ref{fig:FAKT.png}. However, it is worth noting that $M$ is almost equal to one for the main cases of interest (Copper, Aluminium, Stainless Steel 316L etc.). It starts to have a non-negligible impact for the extreme case of very poor conductivity and it decreases with frequency. For electrical conductivity of $\SI{1.67e5}{S/m}$ $M \approx 1$ for frequencies below $\SI{2}{GHz}$. Consequently, in many practical cases, one can ignore the factor $M$ and use the formula in its simplest and most useful formulation of Eq.~(\ref{eq}), with $K=1/2\pi$. Figure~\ref{fig:BeyondThickWall.png} shows the agreement between the impedance from the theory of Eq.~(\ref{Theory}) and the impedance from the analytical derivation. A perfect agreement is observed in the whole frequency range, even beyond the thick wall regime.

\section{Simulation results and comparison with theory}
\label{Comp}
\begin{figure}[!htb]
   \centering
   \includegraphics*[width=.82\columnwidth]{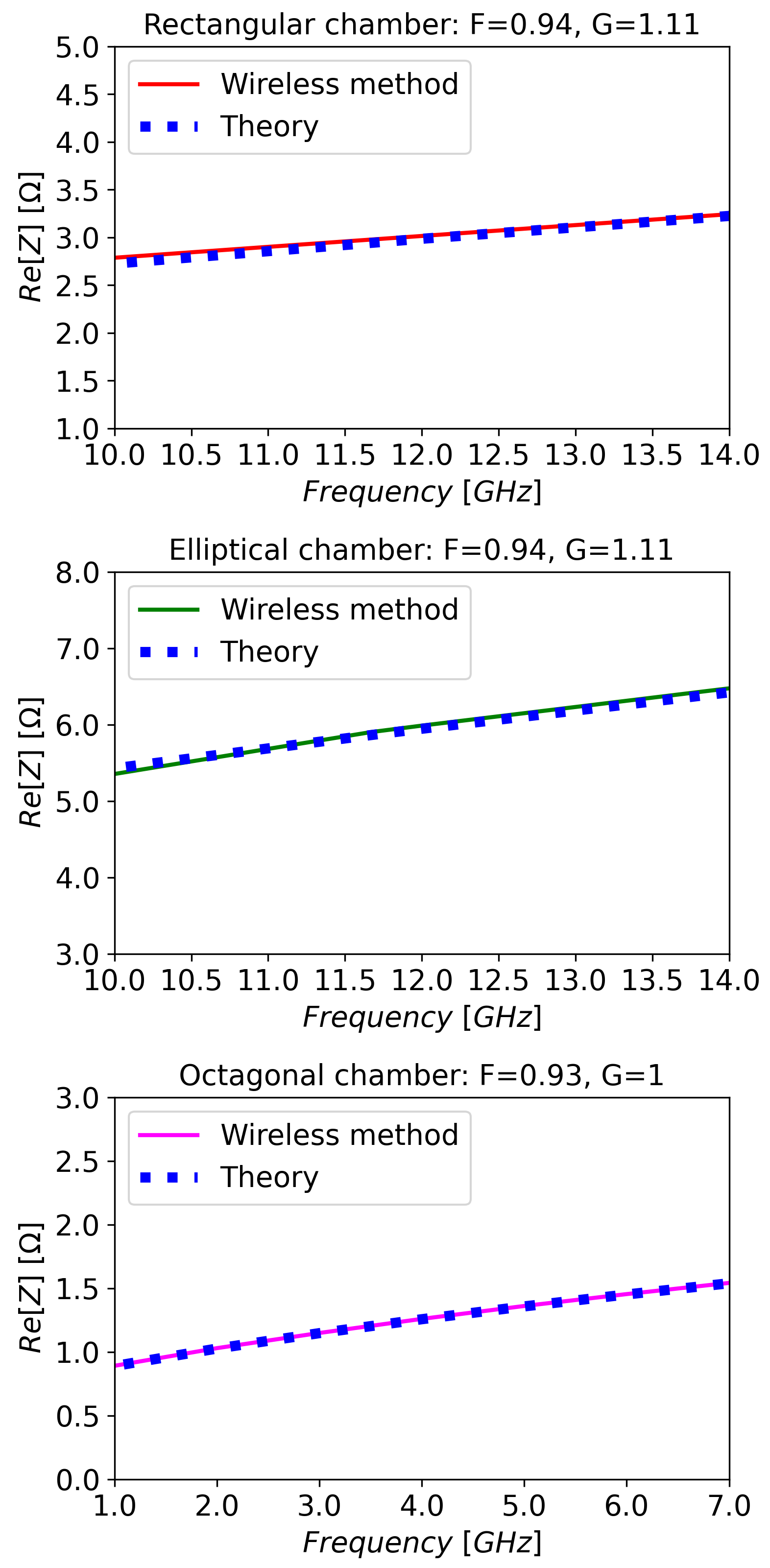}
   \caption{Comparison between the longitudinal impedance obtained from CST simulations (wireless method) and the theoretical impedance for various shapes of the chamber, above their cut-off frequencies. At the top: rectangular chamber with $a=\SI{1.5}{mm}$, $b=\SI{10}{mm}$, in the middle elliptical chamber with $a=\SI{1.5}{mm}$, $b=\SI{10}{mm}$, at the bottom octagonal chamber with $a=b=\SI{50}{mm}$. The imaginary part is not displayed since, in the frequency range analyzed (thick wall regime), it is exactly equal to the real part (see Eq.~(\ref{Theory})).}
   \label{fig:PlotChambersImp}
\end{figure}

The simulation studies are carried out using a 3D electromagnetic tool, CST Studio Suite, which includes several simulation solvers~\cite{cst}. In this framework, the frequency domain solver was chosen because it is equipped with tetrahedral mesh cells, that allow a better discretization of the domain of interest, contrary to the time domain solver where only hexahedral mesh cells are available. Indeed, the aim of the proposed method is also to establish an accurate procedure to compute the impedance of curved and complex geometries. % To determine the right number of tetrahedrons, the numerical convergence of the simulation results versus the number of tetrahedrons has been studied.
The DUT is excited using the Waveguide Ports which allow only the desired TM mode to be launched.
The longitudinal impedance computed from frequency domain simulations by using Eq.~(\ref{eqGen}), has been compared with the exact theoretical evaluation in Fig.~\ref{fig:PlotChambersImp}.
The results show an almost perfect agreement between the two curves suggesting that the proposed simulation approach, with the related formula, is a suitable and accurate method to obtain the beam coupling impedance of arbitrarily shaped chambers.

\section{Conclusions and outlook}
We identified a logarithmic formula that relates the longitudinal beam coupling impedance and the transmission scattering parameter without modification of the DUT. The new formula has been analytically validated for a resistive circular chamber below and above the cut-off frequency. 

The generalization of the method to arbitrary chamber shapes by means of appropriate factors, such as the form factor $F$ and the geometrical factor $G$, has also been discussed and successfully benchmarked with simulations.

Since the new formula relates the longitudinal beam coupling impedance to the scattering parameter, which is also an output of the measurements, this very promising method could pave the way to develop a bench measurement technique, that will not require modifications of the DUT. As testified in Fig.~\ref{fig:BeyondThickWall.png}, the method could allow to obtain the impedance and, therefore, the electromagnetic properties of wall structures (e.g. coating surfaces like NEG, amorphous carbon or Titanium). Furthermore, it could be suitable for measuring the impedance of complex devices and particularly dominated by the resistive wall impedance, such as collimators, of which their impedance is responsible for a sizeable fraction of the impedance budget of the ~\cite{LHCImpedanceModel}.

The key point of the development of a bench measurement technique is the engineering of an excitation able to emulate as much as possible the ideal excitation of the first TM mode provided by the Waveguide Port.
This stage is currently under development as well as the possible extension of the method to resonant structures.

\section{Appendix}
\label{AppS21}
\subsection{Analytical expression of $S_{21}$ for a circular beam chamber in the general case}
The analytical expression of the $S_{21}$ of the resistive chamber is derived, in the general case, from the propagation constant $k_z=\beta-j\alpha$, by imposing the conservation of the energy. It turns out to be the following:
\begin{center}
\begin{equation}
    S_{21}=|S_{21}|e^{j\Phi l}=e^{\alpha L}e^{j \beta L},
    \label{S21}
\end{equation}
\end{center}
where $\Phi$ is the phase of the scattering parameter, $\alpha$ is the attenuation constant, $\beta$ is the phase constant and $L$ the length of the chamber.

\subsubsection{Computation of the propagation constant beyond the thick wall regime: TM modes}
The attenuation constant $\alpha$ for the resistive circular pipe can be expressed by the following formula:
$$\scalemath{0.7} 
{\alpha_{TM}=\Im{\Big[\sqrt{k_{0}^2-\frac{u^2}{b^2} - \frac{t_{1}}{b^2 D}(2t_{2}u + \frac{t_{1}}{D}(t_{2}^2-t_{3}^2) )-j\frac{2t_{1}}{b^2D}(ut_{1} + \frac{t_{1}}{D}t_{2}t_{3})}}\Big],
}$$

The propagation constant $\beta$ can be expressed by the following formula:
$$\scalemath{0.7} 
{\beta_{TM}=Re[\sqrt{k_{0}^2-\frac{u^2}{b^2} - \frac{t_{1}}{b^2 D}(2t_{2}u + \frac{t_{1}}{D}(t_{2}^2-t_{3}^2) )-j\frac{2t_{1}}{b^2D}(ut_{3} + \frac{t_{1}}{D}t_{2}t_{3})}],
}$$
where:
\begin{itemize}
	\item $t_{1}=\frac{k^2_{0}b}{\omega u}=\frac{\omega\mu_{0}\epsilon_{0}b}{u}$
	\item $t_{2}=-\zeta_{J}(\mu_{0}-\epsilon_{0}(\zeta^2_{R}+\zeta^2_{J}))=-\zeta_J N_2$
	\item $t_{3}=\zeta_{R}(\mu_{0}+\epsilon_{0}(\zeta^2_{R}+\zeta^2_{J}))=\zeta_R N_1$
	\item $D=\mu^2_{0}+\epsilon^2_{0}(\zeta^2_{R}+\zeta^2_{J})^2+2\epsilon_{0}\mu_{0}(\zeta^2_{R}-\zeta^2_{J})$
\end{itemize}

It is necessary to derive the square root of the following:
$$\scalemath{0.8}{
\sqrt{k_{0}^2-\frac{u^2}{b^2} - \frac{t_{1}}{b^2 D}(2t_{2}u + \frac{t_{1}}{D}(t_{2}^2-t_{3}^2) )-j\frac{2t_{1}}{b^2D}(ut_{3} + \frac{t_{1}}{D}t_{2}t_{3})}},$$
which is a complex term in the form $ Z=A+jB $. Its square root is given by:
\begin{equation}
\sqrt{Z}=\sqrt{|Z|}\cdot(\cos\frac{\theta}{2}+j\sin\frac{\theta}{2}),
\end{equation}
where
\begin{itemize}
	\item $\theta$ is the argument of the complex number: 
	\begin{center}
		%\begin{itemize}
		    $\theta=\arctan{(\frac{B}{A})}, A>0$\\
		    $\theta=\arctan{(\frac{B}{A})} -\pi, A<0 $\\
		%\end{itemize}
	\end{center}
	\item $|Z|$ is the module of the complex number:
	\begin{center}
		$|Z|=\sqrt{A^2+B^2}$.
	\end{center}
    \end{itemize}
Therefore $\alpha$ and $\beta$ are given by:\\

$\alpha=\sqrt{|Z|}\cdot \sin\frac{\theta}{2}$

$\beta=\sqrt{|Z|}\cdot \cos\frac{\theta}{2}$\;.\\

They can be written as:
\begin{center}
	$A=k_{0}^2-\frac{u^2}{b^2} - \frac{t_{1}}{b^2 D}(2t_{2}u + \frac{t_{1}}{D}(t_{2}^2-t_{3}^2) )$\\
	$ $ \\
	$B=-\frac{2t_{1}}{b^2D}(ut_{3} + \frac{t_{1}}{D}t_{2}t_{3})$\;.
\end{center}

It is important to note that $A$ changes in frequency, and in particular, is negative below cut off and positive above, so it is necessary to split the computation in the two cases.
\begin{itemize}
	\item \textbf{Below cut-off}: 
 $$\scalemath{0.7} {A=k_{0}^2-\frac{u^2}{b^2} - \frac{t_{1}}{b^2 D}(2t_{2}u + \frac{t_{1}}{D}(t_{2}^2-t_{3}^2) )<0}$$ and $$\scalemath{0.9} {|A|>>|B|},$$\\
making mathematical rearrangements turns out the following:	$ $ \\
 $$\scalemath{0.7} {\sqrt{|Z|} = \sqrt{k_{0}^2-\frac{u^2}{b^2}-\frac{t_{1}}{Db^2}(2t_{2}u+\frac{t_{1}}{D}(t_{2}^2-t_{3}^2))}}$$\\
$$\scalemath{0.7} {\sqrt{|Z|} = \sqrt{k_{0}^2-\frac{u^2}{b^2}+ \frac{2\omega\epsilon_0}{b\mu_0}\zeta_J (\mu_0-\epsilon_0 (\zeta^2_R+\zeta^2_J))}}$$\\
$$\scalemath{0.7} {\sqrt{|Z|}\approx \sqrt{(k_{0}^2-\frac{u^2}{b^2})(1+ \frac{\frac{\omega\epsilon_{0}}{b\mu_{0}}2\zeta_{J}(\mu_{0}-\epsilon_{0}(\zeta^2_{R}+\zeta^2_{J}))}{k_{0}^2-\frac{u^2}{b^2}})}}$$\\
considering that:\\
$$\scalemath{0.8} { \delta= \frac{\frac{\omega\epsilon_{0}}{b\mu_{0}}2\zeta_{J}(\mu_{0}-\epsilon_{0}(\zeta^2_{R}+\zeta^2_{J}))}{k_{0}^2-\frac{u^2}{b^2}} << 1  } $$\\ it can be expanded in McLaurin series restricted at the first order, as follows:\\
%$$\scalemath{0.7} {\sqrt{1+\frac{\frac{\omega\epsilon_{0}}{r\mu_{0}}2\zeta_{J}(\mu_{0}-\epsilon_{0}(\zeta^2_{R}+\zeta^2_{J}))}{k_{0}^2-\frac{u^2}{r^2}}}=1+\frac{1}{2} \frac{\frac{\omega\epsilon_{0}}{r\mu_{0}}2\zeta_{J}(\mu_{0}-\epsilon_{0}(\zeta^2_{R}+\zeta^2_{J}))}{k_{0}^2-\frac{u^2}{r^2}} }$$\\
%$$\scalemath{0.7} {\sqrt{1+\frac{\frac{\omega\epsilon_{0}}{r\mu_{0}}2\zeta_{J}(\mu_{0}-\epsilon_{0}(\zeta^2_{R}+\zeta^2_{J}))}{k_{0}^2-\frac{u^2}{r^2}}} \approx 1+\frac{1}{2} \frac{\frac{\omega\epsilon_{0}}{r\mu_{0}}2\zeta_{J}(\mu_{0}-\epsilon_{0}(\zeta^2_{R}+\zeta^2_{J}))}{k_{0}^2-\frac{u^2}{r^2}} }$$\\
$$\scalemath{0.7} {\sqrt{1+\delta} \approx 1+\frac{\delta}{2} }$$\\
$$\scalemath{0.7} {\sqrt{|Z|}\approx \sqrt{k_{0}^2-\frac{u^2}{b^2}} (1+\frac{\frac{\omega\epsilon_{0}}{b\mu_{0}}\zeta_{J}(\mu_{0}-\epsilon_{0}(\zeta^2_{R}+\zeta^2_{J}))}{k_{0}^2-\frac{u^2}{b^2}}) }$$\\
$$\scalemath{0.7} {\theta = \arctan(\gamma) -\pi \approx \gamma -\pi},$$\\

where $\gamma=\frac{B}{A}=-\zeta_R \frac{2 N_1 \omega \epsilon_0 \mu_0}{ ub D}\frac{(u+\frac{t_{1} t_{2}}{D})}{(k_{0}^2-\frac{u^2}{b^2})} << 1.$ \\

Therefore $\alpha$ and $\beta$ can be written:\\

$$\scalemath{0.8} { \alpha\approx\sqrt{k_{0}^2-\frac{u^2}{b^2}} (1+\frac{\frac{\omega\epsilon_{0}}{b\mu_{0}}\zeta_{J}(\mu_{0}-\epsilon_{0}(\zeta^2_{R}+\zeta^2_{J}))}{k_{0}^2-\frac{u^2}{b^2}})\sin{(\frac{\gamma}{2}-\frac{\pi}{2})} }$$\\
$$\scalemath{0.8} { \beta\approx\sqrt{k_{0}^2-\frac{u^2}{b^2}} (1+\frac{\frac{\omega\epsilon_{0}}{b\mu_{0}}\zeta_{J}(\mu_{0}-\epsilon_{0}(\zeta^2_{R}+\zeta^2_{J}))}{k_{0}^2-\frac{u^2}{b^2}})\cos{(\frac{\gamma}{2}-\frac{\pi}{2})} }$$\\
$$\scalemath{0.8} { \beta \approx \sqrt{k_{0}^2-\frac{u^2}{b^2}} (1+\frac{\frac{\omega\epsilon_{0}}{b\mu_{0}}\zeta_{J}(\mu_{0}-\epsilon_{0}(\zeta^2_{R}+\zeta^2_{J}))}{k_{0}^2-\frac{u^2}{b^2}})\sin{\frac{\gamma}{2}}, }$$\\

considering $\gamma << 1$, $\sin\gamma \approx \gamma$ therefore:\\

$$\scalemath{0.8} { \beta \approx \sqrt{k_{0}^2-\frac{u^2}{b^2}} (1+\frac{\frac{\omega\epsilon_{0}}{b\mu_{0}}\zeta_{J}(\mu_{0}-\epsilon_{0}(\zeta^2_{R}+\zeta^2_{J}))}{k_{0}^2-\frac{u^2}{b^2}})\frac{\gamma}{2} }$$ \\
%$$\scalemath{0.8} { \beta \approx - \sqrt{k_{0}^2-\frac{u^2}{b^2}}\gamma }$$ \\
$$\scalemath{0.8} { \beta \approx - \sqrt{k_{0}^2-\frac{u^2}{b^2}}\zeta_R \frac{N_1 \omega \epsilon_0 \mu_0}{ ub D}\frac{(u+\frac{t_{1}t_{2}}{D})}{(k_{0}^2-\frac{u^2}{b^2})} }$$ \\
$$\scalemath{0.8} { \beta \approx -  \zeta_R \frac{N_1 \omega \epsilon_0 \mu_0}{ ub D}\frac{(u+\frac{t_{1} t_{2}}{D})}{\sqrt{(k_{0}^2-\frac{u^2}{b^2})}} }$$ \\
In particular:
\begin{equation}
\scalemath{0.7} { \alpha_{DUT}\approx -\sqrt{k_{0}^2-\frac{u^2}{b^2}} (1+\frac{\frac{\omega\epsilon_{0}}{b\mu_{0}}\zeta_{J}(\mu_{0}-\epsilon_{0}(\zeta^2_{R}+\zeta^2_{J}))}{k_{0}^2-\frac{u^2}{b^2}}) }
\label{alphaDBe}
\end{equation}
\begin{equation}
\scalemath{0.8} { \alpha_{REF} \approx -\sqrt{k_{0}^2-\frac{u^2}{b^2}} }
\label{alphaRBe}
\end{equation}
\begin{equation}
\scalemath{0.8} { \beta_{DUT} \approx -  \zeta_R \frac{N_1 \omega \epsilon_0 \mu_0}{ ub D}\frac{(u+\frac{t_{1} t_{2}}{D})}{\sqrt{(k_{0}^2-\frac{u^2}{b^2})}} }
\label{betaDBe}
\end{equation}
\begin{equation}
\scalemath{0.8} { \beta_{REF} \approx 0 }
\label{betaRBe}
\end{equation}

\item \textbf{Above cut-off}: 
$$A=\scalemath{0.8} { k_{0}^2-\frac{u^2}{r^2} - \frac{t_{1}}{r^2 D}(2t_{2}u + \frac{t_{1}}{D}(t_{2}^2-t_{3}^2) )>0 } $$\\ and $$\scalemath{0.9} {|A|>>|B|},$$\\
making mathematical rearrangements turns out the following:	$ $ \\
%$$\scalemath{0.7} {\sqrt{|Z|} = \sqrt{k_{0}^2-\frac{u^2}{b^2}-\frac{t_{1}}{Dr^2}(2t_{2}u+\frac{t_{1}}{D}(t_{2}^2-t_{3}^2))}}$$\\
$$\scalemath{0.7} {\sqrt{|Z|} = \sqrt{k_{0}^2-\frac{u^2}{b^2}+ \frac{2\omega\epsilon_0}{b\mu_0}\zeta_J (\mu_0-\epsilon_0 (\zeta^2_R+\zeta^2_J))}}$$\\
$$\scalemath{0.7} {\sqrt{|Z|} = \sqrt{(k_{0}^2-\frac{u^2}{b^2})(1+ \frac{\frac{\omega\epsilon_{0}}{b\mu_{0}}2\zeta_{J}(\mu_{0}-\epsilon_{0}(\zeta^2_{R}+\zeta^2_{J}))}{k_{0}^2-\frac{u^2}{b^2}})}}$$\\

Therefore $\alpha$ and $\beta$ can be written:\\

$$\scalemath{0.8} { \alpha\approx\sqrt{(k_{0}^2-\frac{u^2}{b^2})(1+ \frac{\frac{\omega\epsilon_{0}}{b\mu_{0}}2\zeta_{J}(\mu_{0}-\epsilon_{0}(\zeta^2_{R}+\zeta^2_{J}))}{k_{0}^2-\frac{u^2}{b^2}})} \sin{\frac{\gamma}{2}} }$$\\
$$\scalemath{0.8} { \beta\approx\sqrt{(k_{0}^2-\frac{u^2}{r^2})(1+ \frac{\frac{\omega\epsilon_{0}}{r\mu_{0}}2\zeta_{J}(\mu_{0}-\epsilon_{0}(\zeta^2_{R}+\zeta^2_{J}))}{k_{0}^2-\frac{u^2}{b^2}})} \cos{\frac{\gamma}{2}}} ,$$\\

considering $\gamma << 1$, $\sin\gamma \approx \gamma$ therefore:\\

$$\scalemath{0.8} { \alpha \approx \sqrt{k_{0}^2-\frac{u^2}{b^2}} (1+\frac{\frac{\omega\epsilon_{0}}{b\mu_{0}}\zeta_{J}(\mu_{0}-\epsilon_{0}(\zeta^2_{R}+\zeta^2_{J}))}{k_{0}^2-\frac{u^2}{b^2}})\frac{\gamma}{2} }$$ \\
$$\scalemath{0.8} { \beta \approx \sqrt{k_{0}^2-\frac{u^2}{b^2}} (1+\frac{\frac{\omega\epsilon_{0}}{b\mu_{0}}\zeta_{J}(\mu_{0}-\epsilon_{0}(\zeta^2_{R}+\zeta^2_{J}))}{k_{0}^2-\frac{u^2}{b^2}}) }$$ \\
%$$\scalemath{0.8} { \beta \approx - \sqrt{k_{0}^2-\frac{u^2}{r^2}}\frac{\gamma}{2} }$$ \\
$$\scalemath{0.8} { \alpha \approx  \sqrt{k_{0}^2-\frac{u^2}{b^2}}  \zeta_R \frac{t_{3}^* \omega \epsilon_0 \mu_0}{ ub D}\frac{(u+\frac{t_{1} t_{2}}{D})}{{(k_{0}^2-\frac{u^2}{b^2})}} }$$ 

In particular:
\begin{equation}
\scalemath{0.8}{\alpha_{DUT}\approx -  \zeta_R \frac{t_{3}^* \omega \epsilon_0 \mu_0}{ ub D}\frac{(u+\frac{t_{1} t_{2}}{D})}{\sqrt{(k_{0}^2-\frac{u^2}{b^2})}} }
\label{alphaDAb}
\end{equation}
\begin{equation}
\scalemath{0.8}{\alpha_{REF}\approx 0 }
\label{alphaRAb}
\end{equation}
\begin{equation}
\scalemath{0.7} {\beta_{DUT} \approx -  \sqrt{k_{0}^2-\frac{u^2}{b^2}} (1+\frac{\frac{\omega\epsilon_{0}}{b\mu_{0}}\zeta_{J}(\mu_{0}-\epsilon_{0}(\zeta^2_{R}+\zeta^2_{J}))}{k_{0}^2-\frac{u^2}{b^2}})}
\label{betaDAb}
\end{equation}
\begin{equation}
\scalemath{0.8}{\beta_{REF} \approx -\sqrt{k_{0}^2-\frac{u^2}{b^2}}}
\label{betaRAb}
\end{equation}

\end{itemize}

\section{Analytical validation of the wireless log-Formula for a circular beam chamber in the general case}
\label{AppValidation}
Considering the proposed equation to determine the impedance:
    \begin{equation}
    Z= -K \cdot Z_{TM} \ln{\frac{S^{DUT}_{21}}{S^{REF}_{21}}}.
    \label{eqCir}
    \end{equation}
in this section is shown that it reduces to Eq.~\ref{Theory} by substituting Eq.~\ref{S21} and finding the proper expression for the $K$ term. The mode impedance $Z_{TM}$ is imaginary below cut-off and real above cut-off.
\begin{itemize}
    \item Below cut-off \\
    Substituting Eq.~\ref{S21} in Eq.~\ref{eqCir} :\\
    $$\scalemath{0.8} {Z=-K Z_{TM}\cdot((\alpha_{DUT}-\alpha_{REF})L+j(\beta_{DUT}-\beta_{REF})L)} $$ \\
    
    in particular:
    
    \begin{equation}
    \scalemath{0.8} {\Re{(Z)}=-K |Z_{TM}|\cdot(\beta_{DUT}-\beta_{REF}))L}
    \label{ReBeZ}
    \end{equation}
    \begin{equation}
    \scalemath{0.8} {\Im{(Z)}=-K |Z_{TM}|\cdot(\alpha_{DUT}-\alpha_{REF})L}
    \label{ImBeZ}
    \end{equation}
    
    and substituting Eqs.~(\ref{alphaDBe}),(\ref{alphaRBe}),(\ref{betaDBe}),(\ref{betaRBe}) in Eqs.(\ref{ReBeZ}),(\ref{ImBeZ}) :
    
    \begin{equation}
    \scalemath{0.6} {\Re{(Z)}=-K \frac{\sqrt{k^2_0-\frac{u^2}{b^2}}}{\omega\epsilon_0}\cdot\frac{\zeta_R}{b}\frac{N_1\mu_0}{D u}(u+\frac{t_{1} t_{2}}{D})\frac{\omega\epsilon_0}{\sqrt{k^2_0-\frac{u^2}{b^2}}} L =}
    \label{ReBeZfinal}
    \end{equation}
    $$ \scalemath{0.6} {= K\frac{\zeta_R L}{ b} M }$$\\
    
    where:
    
    $$ \scalemath{0.8} {M= \frac{N_1 \mu_0}{D u}(u-\frac{\omega\mu_{0}\epsilon_{0}b}{u} \frac{\zeta_{J}N_{2}}{D} ) } $$ \\

    \begin{equation}
    \scalemath{0.7} {\Im{(Z)}=K\frac{{k^2_0-\frac{u^2}{b^2}}}{\omega\epsilon_0}\cdot\ \frac{\frac{\omega\epsilon_{0}}{b\mu_{0}}\zeta_{J}(\mu_{0}-\epsilon_{0}(\zeta^2_{R}+\zeta^2_{J}))}{k_{0}^2-\frac{u^2}{b^2}} L = }
    \label{ImBeZfinal}
    \end{equation}
    $$ \scalemath{0.7} {= K\frac{\zeta_J L}{b} (\frac{\mu_0-\epsilon_0 (\zeta^2_R+\zeta^2_J)}{\mu_0})  \approx K \frac{\zeta_J L}{ b} } $$ \\

    \item Above cut-off \\
    Substituting Eq. \ref{S21} in Eq. \ref{eqCir} :\\
    $$\scalemath{0.8} {Z=-K Z_{TM}\cdot((\alpha_{DUT}-\alpha_{REF})L+j\beta_{DUT}L)} $$ \\
    
    in particular:
    
    \begin{equation}
    \scalemath{0.8} {\Re{(Z)}=-K |Z_{TM}|\cdot(\alpha_{DUT}-\alpha_{REF})L}
    \label{ReAbZ}
    \end{equation}
    \begin{equation}
    \scalemath{0.8} {\Im{(Z)}=-K |Z_{TM}|\cdot(\beta_{DUT}-\beta_{REF})L}
    \label{ImAbZ}
    \end{equation}
    
    and substituting Eqs.~(\ref{alphaDAb}),(\ref{alphaRAb}),(\ref{betaDAb}),(\ref{betaRAb}) in Eqs.~(\ref{ReAbZ}),(\ref{ImAbZ}) :
    
    \begin{equation}
    \scalemath{0.6} {\Re{(Z)}=K\frac{\sqrt{k^2_0-\frac{u^2}{b^2}}}{\omega\epsilon_0}\cdot\frac{\zeta_R}{b}\frac{N_1\mu_0}{D u}(u+\frac{t_{1} t_{2}}{D})\frac{\omega\epsilon_0}{\sqrt{k^2_0-\frac{u^2}{b^2}}} L =}
    \label{ReAbZfinal}
    \end{equation}
    $$ \scalemath{0.6} {= K\frac{\zeta_R L}{b}\frac{N_1 \mu_0}{Du}(u+\frac{t_{1} t_{2}}{D})=K\frac{\zeta_R L}{ b} M} $$\\
    
    where:
    
    $$ \scalemath{0.8} {M= \frac{N_1 \mu_0}{D u}(u-\frac{\omega\mu_{0}\epsilon_{0}b}{u} \frac{\zeta_{J}N_{2}}{D} ) }$$ \\
    
    \begin{equation}
    \scalemath{0.7} {\Im{(Z)}=K\frac{{k^2_0-\frac{u^2}{b^2}}}{\omega\epsilon_0}\cdot\ \frac{\frac{\omega\epsilon_{0}}{b\mu_{0}}\zeta_{J}(\mu_{0}-\epsilon_{0}(\zeta^2_{R}+\zeta^2_{J}))}{k_{0}^2-\frac{u^2}{b^2}} L = }
    \label{ImAbZfinal}
    \end{equation}
    $$ \scalemath{0.7} {= K \frac{\zeta_J L}{b} (\frac{\mu_0-\epsilon_0 (\zeta^2_R+\zeta^2_J)}{\mu_0})  \approx K \frac{\zeta_J L}{ b} } $$ \\

    It is clear that Eq.~(\ref{ReBeZfinal}) and Eq.~(\ref{ReAbZfinal}) reduces to Eq.~(\ref{Theory}) if $K=\frac{1}{2\pi M} $ and that Eq.~(\ref{ImBeZfinal}) and Eq.~(\ref{ImAbZfinal}) reduces to Eq.~(\ref{Theory}) if $K=\frac{1}{2\pi}$ in Eq.~(\ref{eqCir}).
    %since the $(\mu_{0}-\epsilon_{0}(\zeta^2_R+\zeta^2_J){\mu_0})\approx 1 $.

\end{itemize}

\printbibliography

%\vspace{20cm}
\end{document}